\documentclass{desyproc}

\begin{document}
\title{Searches for very weakly-coupled particles beyond the Standard Model with NA62}

\author{{\slshape 
Babette D\"obrich$^1$, for the NA62 collaboration
\footnote{
The NA62 Collaboration: R.~Aliberti, F.~Ambrosino, R.~Ammendola, B.~Angelucci, A.~Antonelli, G.~Anzivino, R.~Arcidiacono, 
M.~Barbanera, A.~Biagioni, L.~Bician
% ...
}} \\
$^1$CERN, 1211 Geneva, Switzerland\\
}

\contribID{dobrich\_babette}

\confID{13889}  
\desyproc{DESY-PROC-2017-XX}
\acronym{Patras 2017} 
\doi  

\maketitle

\begin{abstract}
The NA62 experiment at CERN is designed 
to measure precisely the rare decay $K^{+} \rightarrow \pi^+  \nu \bar{\nu}$.
The intensity and energy of the SPS proton beam used to produce the $K^+$,
as well as the hermetic detector coverage and overall geometry, give in addition
the opportunity to
search for hypothesized weakly-coupled particles at the MeV-GeV mass scale.
In these proceedings the focus lies on reviewing these opportunities and sketching the
current status of some pertinent searches.
\end{abstract}

\section{Kaon physics with NA62}

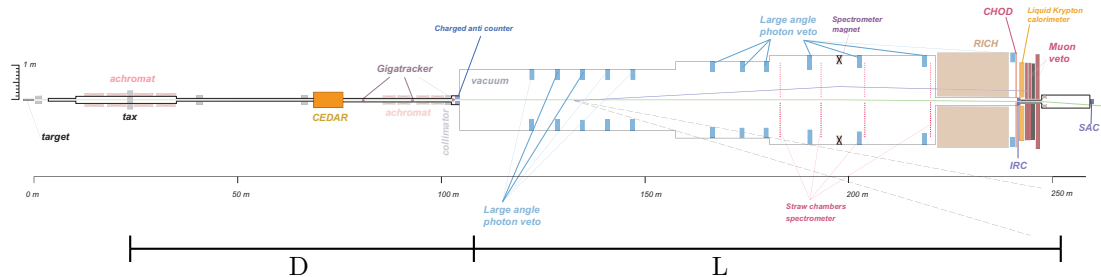
\begin{figure}
\begin{picture}(300,100)(0,-20)
\thicklines
\put(45,-5){\line(1,0){130}}
\put(45,-5){\line(0,1){5}}
\put(45,-5){\line(0,-1){5}}
\put(175,-5){\line(0,1){5}}
\put(175,-5){\line(0,-1){5}}
\put(105,-15){$$D$$}
\put(175,-5){\line(1,0){222}}
\put(397,-5){\line(0,1){5}}
\put(397,-5){\line(0,-1){5}}
\put(265,-15){$$L$$}
 \includegraphics[width=1\textwidth]{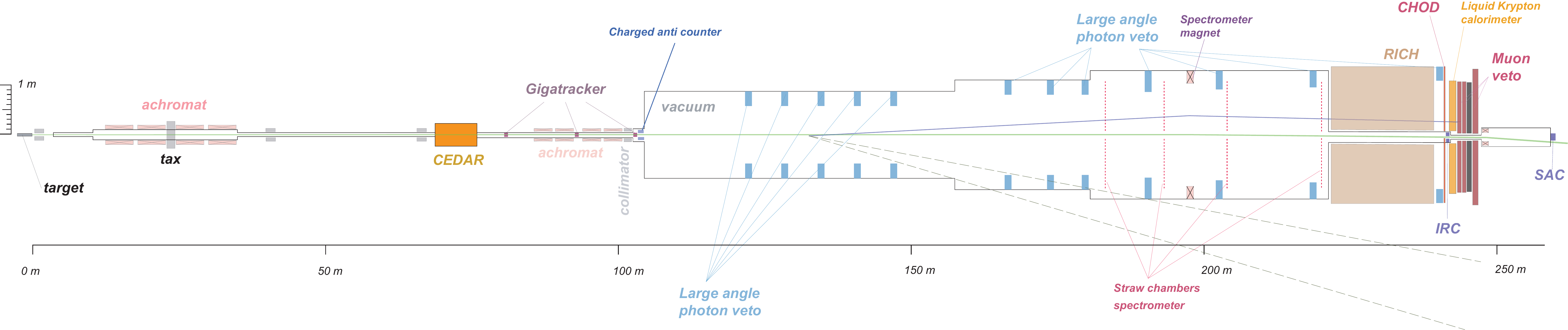}
\end{picture}
 \caption{Layout of NA62: 
 The SPS proton beam (from left) hits a beryllium target.
 A fraction of the secondary beam as well as
 around $\sim$40\% of the primary protons that
 have not interacted in the target are guided towards an achromat,
 located
 approximately 24 m downstream of the target.
 This achromat, called `TAX', is
  described in more detail in Figure \ref{Fig:Tax}.
 All beam components
 (including the remaining 400 GeV component)
 except for a positive component at $\sim$75 GeV, are stopped in the TAX.
 The 75 GeV component eventually is guided along a distance $D$
 to the start of a decay volume of length $L$ downstream.
Kaons from the secondary hadron beam are eventually identified (CEDAR) and measured (Gigatracker).
}
 \label{fig:na62}
\end{figure}

The NA62 experiment
aims to measure to a good accuracy the very small branching ratio (BR) of $K^+ \rightarrow \pi^{+} \nu \bar{\nu}$,
which is $\mathcal{O}(10^{-10})$.
This BR is rather precisely ($<$10 \% level) predicted from theory and has a recognized 
sensitivity to new physics \cite{Buras:2015yca}.
However,
the only existing measurement of this BR is based on only seven events \cite{Artamonov:2009sz},
and the experimental precision is thus not sufficient to challenge the theory prediction.
NA62 aims to measure this BR at the 10\% level by combining a large number of kaon  decays $\mathcal{O}(10^{13})$
 with a hermetic detector system,
that can reject other kaon decays in the $\sim60$ m long fiducial volume (FV)
ultimately at the $\mathcal{O}(10^{12})$ level.
The detector layout is shown in Fig.~\ref{fig:na62},
and a detailed description of the physics case, the experiment and its performance  is provided in
\cite{NA62:2017rwk}. 

To date, NA62
has released a preliminary analysis of 5\% of 2016 data, results are described in \cite{spsc}.
In 2017, $\sim3\times 10^{12}$ kaon decays have been collected and
the experiment will continue data taking in 2018, until the long shutdown of the CERN accelerators.

Besides the main measurement, NA62 has a program to search for 
lepton number and lepton flavor violating decays.
In addition, with existing and near-future data,
NA62 has a discovery potential in the search of very weakly-coupled 
particles beyond the Standard Model. 
Such particles, if existing at the MeV-GeV
scale would have comparably long lifetimes
and could have escaped all past detection efforts so far.
The need for an increased experimental activity
in this area is widely appreciated in the HEP community, since the existence of a `Dark Sector'
can be linked to a plethora of open physics questions \cite{Alexander:2016aln}.

\section{Beam-line and production modes of novel particles}

\begin{figure}
\centerline{\includegraphics[width=0.75\textwidth]{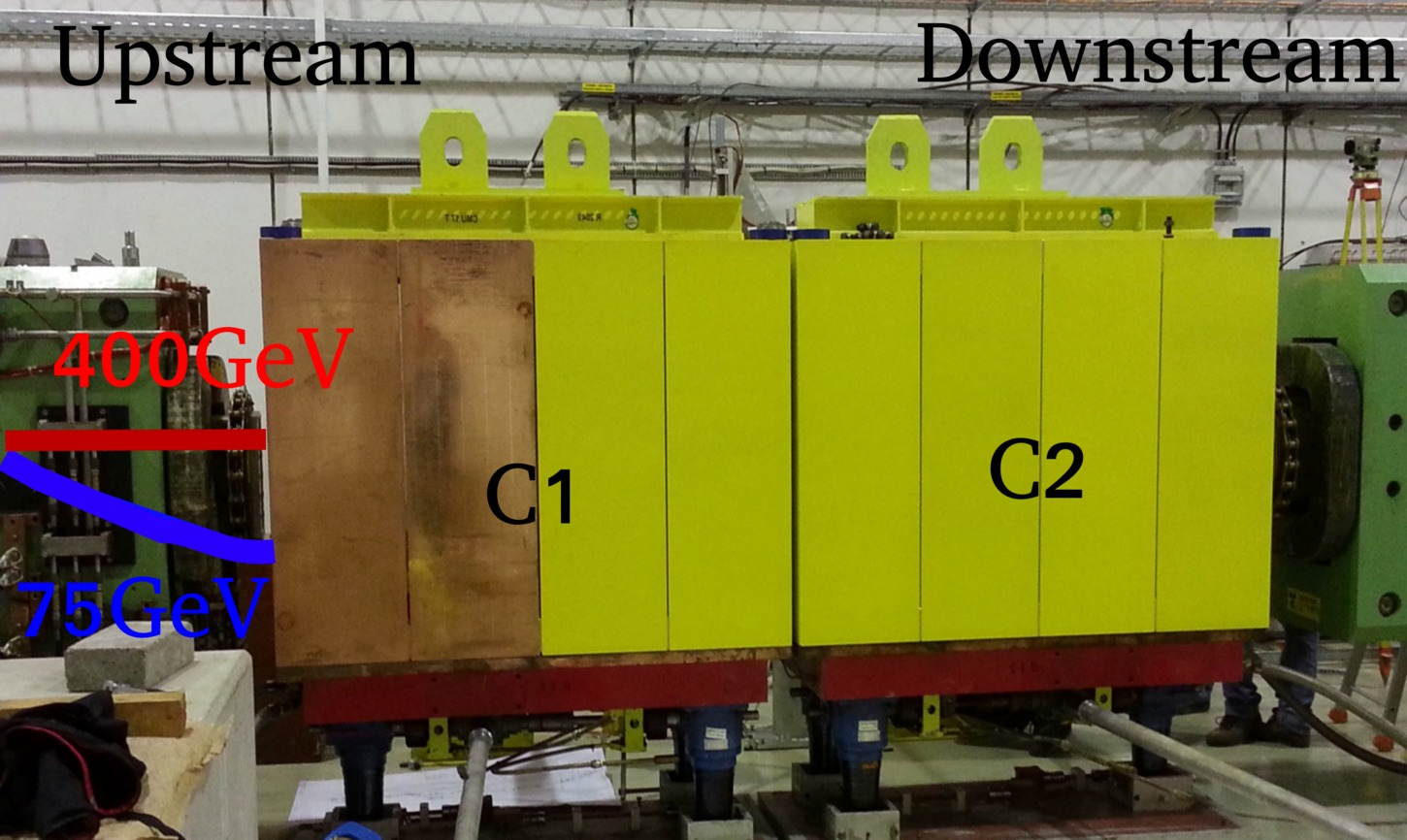}}
\caption{Picture of the `TAX' achromat of the NA62 beamline:
It is located $\sim$80 m upstream of the NA62 decay volume.
The beam is incident from the left 
and is dispersed by a magnet before reaching the two collimators, labeled as C1 and C2.
For simplicity, only two positive beam components are sketched, whilst in reality a wide momentum spectrum
reaches the collimators.
C1 (copper/iron) and C2 (full iron) are each $\sim 1.6$ m long and
each contain small bores (from left to right, not visible in the picture)
to allow the passage
of the $75$ GeV component (sketched in blue) in nominal data taking.
The two collimator blocks can be moved along the vertical direction, to
either reduce the intensity or completely block the beam (by misaligning the bores) and act as a beam-dump.
The upstream beryllium target (not visible in the picture)
can be 
removed for special runs. In this situation the entire 400 GeV primary proton component
is dumped into C1.
}\label{Fig:Tax}
\end{figure}

The kaons of interest for NA62 are produced by interaction of the
SPS 400 GeV proton beam in an upstream target, see Fig.~\ref{fig:na62}.
The unseparated, positively charged hadron beam is selected in momentum
by an achromat, shown and described in Figure \ref{Fig:Tax}.

Consequently, in {\it standard data-taking}, besides beam halo muons, only these secondary  particles (essentially $\pi^+, p, K^+$)
at 75 GeV will enter the decay volume of NA62. 
Interestingly, even during data-taking with the upstream target in place, about 40 \% of the primary protons 
punch through this target and eventually
interact in the first collimator C1, cf. Fig~\ref{Fig:Tax}, with their hadronic shower being absorbed.
{\it Another possible mode for data-taking} in NA62 is to run as a `beam-dump'
such that up to 100\% of the 400 GeV protons are stopped
in the collimators.

 Interactions of 400 GeV protons
in the collimator,  as well as decays of secondary mesons before the
collimator may be the source of different, so-far
undiscovered, weakly-interacting particles,
which travel unhindered to the NA62 decay volume and decay into visible final states.
Besides the primary beam energy
and protons on target (POT),
the main variable for the parameter-space-coverage in these searches is
geometry: Ideally, the decay volume is close enough to novel particle's
production point to contain the decay products of the most short-lived particles
as well as long enough to detect the most long-lived candidates.
The comparably small $D \simeq 80$m, and large $L \simeq 136$ m  (cf. Fig.~\ref{fig:na62})
of the NA62 setup enable 
the experiment to extend search parameters beyond 
past results, e.g. from CHARM and NuCal for axion-like particle searches \cite{Dobrich:2015jyk}.

In both, nominal and dump data-taking modes, competitive searches 
for weakly interacting particles beyond the Standard Model are possible.
NA62 can make a strong contribution in  shedding light
on the existence of `Dark Sector particles' using:

\begin{enumerate}
 \item {\bf Meson decays in the FV:} 
 The large flux of kaons and
 pions in NA62 suggests to search novel particles
 in their decays.
 Examples of recent preliminary results are: Firstly production searches
 of Heavy Neutral Leptons $N$ in $K^+ \rightarrow l^{+} \ N$, with $l= e,\mu$.
 Secondly: the search for invisibly
 decaying Dark Photons $A'$ from $K^+ \rightarrow \pi^+ \pi^0$, where $\pi^0 \rightarrow \gamma \ A'$.
 Preliminary results of both analyses are given in \cite{spsc}.
 \item {\bf Parasitic dump production:}
 The decay of hypothesized new particles in the NA62 decay volume produced upstream
 can be searched for by dedicated trigger chains (that do not require the presence of a kaon)
 in parallel to the main trigger.
 Such decays can be potentially
 disentangled from hadron beam decays and random track combinations.
 In 2017 dedicated trigger chains have been set up for multi-track events with at least one or two muons to record 
 decays from (pseudo-)scalars (axion-like particles), Dark Photons and Heavy Neutral Leptons.
 \item {\bf Dedicated dump runs:} To achieve the best possible sensitivity 
 in the search for weakly-coupled particles produced upstream,
 it is mandatory to `close' the collimators (i.e. mis-align bores for the hadron beam). 
 In this setting,
 background is minimized.
The current statistics collected in this mode 
is of the order of $\lesssim 10^{16}$ POT. Specific
trigger chains for at least one track or a minimum energy in the  Liquid Krypton electromagnetic calorimeter are 
typically required.
 
\end{enumerate}

\section{Outlook and Summary}

The NA62 experiment has been designed to sustain a high beam-rate, provide full particle identification,
hermetic coverage and very light-material tracking.
Besides the main measurement $K^{+} \rightarrow \pi^+  \nu  \bar{\nu}$,
NA62 has the means and a program to discover weakly-coupled 
particles beyond the Standard Model.

Consequently, for the 2021-2023 operation of the CERN accelerators,
an extended ($\sim$ one-year-long) data-taking in dump-mode is being discussed with the aim to
collect around $\sim 10^{18}$ POT. Reference \cite{Tommaso}
gives physics examples, projections and results on background rejection.
In the meanwhile, data collected in 2016/2017 are used to validate
expectations for backgrounds and  to
perform first analyses possible within the statistics
collected so far.

In summary, the NA62 experiment combines the opportunity for precision Standard Model measurements
with the discovery potential for wide range
of weakly-coupled new physics.

\section*{Acknowledgments}

The author would like to thank the PATRAS 2017 organizers for their continued effort in realizing
fruitful and exciting
workshops in the series.


\begin{footnotesize}

\end{footnotesize}



\end{document}